\documentstyle[preprint,aps,pra,psfig]{revtex}
\tightenlines
 
\begin{document}
\draft

\title{Scattering length of the ground state Mg+Mg collision.}

\author{ E. Tiesinga, S. Kotochigova, and P. S. Julienne}
\address{Atomic Physics Division, NIST, 100 Bureau Drive Stop 8423,
                  Gaithersburg, MD 20899-8423, USA.}

\maketitle
\begin{abstract}  
We have constructed the $X^1\Sigma_g^+$ potential for the collision
between two ground state Mg atoms and analyzed the effect of uncertainties
in the shape of the potential on scattering properties at ultra-cold
temperatures.  This potential reproduces the experimental term values
to 0.2 cm$^{-1}$ and has a scattering length of +1.4(5) nm where the
error is prodominantly due to the uncertainty in the dissociation energy
and the $C_6$ dispersion coefficient. A positive sign of the scattering
length suggests that a Bose-Einstein condensate of ground state Mg atoms
is stable.
\end{abstract}

\date{\today}

\section{Introduction}

It was shown in Ref.~\cite{Sengstock} that Mg atoms can be
magneto-optically trapped.  This offers many possible applications of
cold Mg atoms for ultra high resolution spectroscopy, new frequency
standards and collective quantum effects.  A magnesium clock has already been
developed by Ref.~\cite{Ruschewitz98}. The most abundant isotopes of Mg
have a single electronic ground state without hyperfine interactions,
which opens a road to simpler theoretical modeling of ground state
atomic collisions as compared to the modeling of, say, alkali-metal
atom collisions\cite{Jul97}.

Magnesium is not the only alkaline earth species that can be magnetically
and optically manipulated. Ingenious cooling schemes exist for
Ca\cite{Curtis01} and Sr\cite{Katori99,Katori99b,Ido00,Katori00}.
For Sr Ref.~\cite{Katori99} have nearly reached the quantum
degeneracy regime. Optical clocks based on the ultra-cold
calcium have been constructed for Mg\cite{Ruschewitz98} and
Ca\cite{Riehle98,Riehle99,Oates99,Oates01,Udem01}. The first experimental
photoassociation spectra were reported for calcium\cite{Tiemann00}
while photoassociation spectroscopy for alkaline earth atoms has been
studied theoretically in Ref.~\cite{Machholm01}

The coldest temperatures in dilute atomic gases are obtained by a process
called evaporative cooling. Ground state collisions are crucial for
evaporative cooling.  Elastic collisions during this process lead to
a thermalization of atoms and under the right conditions formation of
a Bose condensate. Elastic collision rates at ultra cold temperatures
can be described by a single parameter, the scattering length {\it a}.
Formation of Bose-Einstein condensates is determined by the nonlinear
coupling parameter in the condensate Schr\"{o}dinger equation, which in
turn depends on the sign and value of the scattering length.

Inelastic collisions, which change the internal state of the atoms, can
eject trapped atoms. Examples of inelastic processes are, for example,
spin-exchange, spin-depolarization, and Penning and associative
ionization, where the first two have been observed in alkali-metal
gases\cite{Jul97} and the latter have been observed in metastable rare gas
samples\cite{YYY}.  For alkaline-earth atoms the electronic ground state
is solely composed of closed shells and is a $^1S_0$ state. Consequently,
no inelastic atom-atom collisions can occur. This opens a pathway to
more efficient evaporative cooling.

In this paper we present our calculation of the ground state Mg$_2$
scattering length and cross-section as a function of the collision
energy, using a potential constructed from high resolution spectra of
the magnesium dimer measured by Balfour and Douglas \cite{Balfour}.
At temperatures below 5 mK ground state Mg collisions are in the
s-wave scattering regime and the {\it l} = 0 phase shift determines the
cross-section and, at zero collision energy, the scattering length. This
phase shift can be found by matching the numerically evaluated scattering
wavefunction of the interaction potential to free scattering wavefunctions
at large internuclear separation {\it R}. Slight changes to the inner
part of the potential can generate significant changes in the phase
shift and the scattering length.

We will discuss several ways to obtain the interaction potential of
ground state Mg$_2$.  Firstly, a Rydberg-Klein-Rees (RKR) potential
curve has been constructed in Ref.  \cite {Balfour} from their
measurement of the rovibrational levels ($\nu$ = 0 - 12, $J$ = 10 - 76)
of ground state Mg$_2$.  Secondly, Vidal and Scheingraber \cite{Vidal}
have reevaluated the molecular constants of Ref.~\cite{Balfour} and
improved upon the RKR analyses by applying a variational procedure
based on the inverted perturbation approach (IPA).  Finally, there
exist a large number of theoretical electronic structure calculations
\cite{Liu,Bertoncini,Muhlhausen,Hay,Stevens,Purvis,Diercksen,Chalasinski,Dyall,Klopper,Tao,Czuchaj}
of the ground state Mg$_2$ potential. This paper briefly describes our
{\it ab~initio} multiconfiguration valence bond (MVB) calculation of
the ground state potential.  Although, as we will show, the theoretical
uncertainty in the shape of the potential is too large to predict
the scattering length we nevertheless compare scattering data for our
{\it ab~initio} potential and the most recently published theoretical
potential by Czuchaj {\it et al.} \cite{Czuchaj} to those for the RKR
and IPA potential. This will give us a feeling for the state of the art
in molecular electronic structure calculations of interacting two
electron atoms.

This paper is set up as follows. In Section~\ref{exp} we describe
the existing experimental data and the potentials that have been
constructed from the data.  Section~\ref{abinitio} presents the
theoretical calculations in the existing literature as well as a new
calculation using the multiconfiguration valence bond method. Section
\ref{lr} describes the long-range behavior of the potentials and connects
this to the short-range potentials obtained in Sections~\ref{exp} and
\ref{abinitio}. Finally, Section~\ref{vib} discusses how well bound
states of the four potentials that have been constructed reproduce
the experimental bound state energies, and determines the scattering
properties for the best of these four potentials.

\section{Experimental data and the ground state potential}\label{exp}

The RKR potential of Ref.~\cite{Balfour} and the IPA potential constructed
in this work from data published in Ref.~\cite{Vidal} are shown in
Fig.~\ref{all_pot}.  For the RKR potential we use a dissociation
energy D$_e$ = 424(5)  cm$^{-1}$ (1 cm$^{-1}$ = 29.9792458 GHz),
defined as the energy difference between the bottom of the potential
and the asymptotic energy.  Reference~\cite{Vidal} did not provide a
tabulated IPA potential but expressed the potential in terms of Dunham
coefficients Y$_{lm}$. We have constructed a potential from the Y$_{lm}$
for {\it l} = 0-3, $m$ = 0-5 provided in Table IV of Ref.~\cite{Vidal}
by applying the RKR procedure.  We will call this potential the IPA
potential.  We will show lateron by solving the Schr\"odinger equation
for the RKR and IPA potentials that the IPA potential reproduces the term
values significantly better than the RKR potential of Ref.~\cite{Balfour}.
For the IPA potential we take the dissociation energy D$_e$ to be
431.0(1.0) cm$^{-1}$ in accordance with fit of Ref.~\cite{Vidal} to the
last outer turning points of the potential. The 1.0 cm$^{-1}$ uncertainty
is based on the sensitivity of the fit with the order of the long-range
dispersion expansion.  The RKR and IPA potential are only known over
a limited region of internuclear separation, 6 $a_0$ to 14
$a_0$. This range is determined by the inner and outer turning point of
the most-weakly-bound measured rovibrational level.

For a scattering calculation the potential must be known for all
internuclear separations.  Therefore we have connected the repulsive short
range of the RKR potential to the repulsive wall of our MVB potential
which will be discussed in Section III. The repulsive short range wall
of the IPA potential is a linear extrapolation from the attractive
region as our MVB potential could not be smoothly connected to the IPA
potential. Both RKR and IPA potentials suffer from a well known ``short
range turnover'' in the potential inherent to the RKR inversion procedure.
We simply removed those inner turning points from the data set before
extrapolation.  The extrapolation of the RKR and IPA potential to longer
$R$ is discussed in Section IV.

\section{The ab initio ground state potentials}\label{abinitio}

The ground state Mg$_2$ molecule is formed from two closed shell atoms,
each of which is described by the configuration 1s$^2$2s$^2$2p$^6$3s$^2$.
This might suggest that theoretical modelling will be easy.  Instead,
numerous computational efforts have proven the opposite. The
first dramatic complication arises at the Hartree-Fock level,
because it predicts a purely repulsive  ground state potential.
The binding of the magnesium dimer is created by the correlation
energy only.  There are inter- and intrashell correlations affecting the
potential, that, apparently, have a strong dependence on internuclear
separation.  The question becomes how well a computational approach
can incorporate these correlations.  There are a large number of methods
\cite{Liu,Bertoncini,Muhlhausen,Hay,Stevens,Purvis,Diercksen,Chalasinski,Dyall,Klopper,Tao,Czuchaj}
which have been tested for the computation of the correlation corrections.
In a pioneering publication Stevens and Krauss \cite{Stevens}
calculated the ground state potential of Mg$_2$ using a nonrelativistic
multiconfiguration self-consistent field method.  Their work shows
the importance of the ability to simultaneously incorporate both the
long-range atomic and short-range molecular correlations.  Many-body
perturbation theory can provide an alternative to the configuration
interaction approach to model correlation effects in the Mg$_2$ molecule.
Double excitation type diagrams were applied by Purvis and Bartlett
\cite{Purvis} to significantly improve the molecular binding energy.

In Ref. \cite{Czuchaj} a combination of a coupled-cluster method with
single and double excitations and perturbative triple excitations is used
to reach a good agreement with the RKR potential of Ref. \cite{Balfour}.
The potential curve of Ref. \cite{Czuchaj} is shown in Fig.~\ref{all_pot}.

For this paper we have used the multiconfiguration valence bond
(MVB) method to calculate the ground state Mg$_2$ potential. A
detailed description of the computational approach is given in
Ref.~\cite{Kotochigova}.  We create a nonorthogonal basis set from
self-consistent Dirac-Fock atomic orbitals belonging to the [1s$^2$]
2s$^2$ 2p$^6$ 3s$^2$ configuration and additional Sturmian orbitals
labeled 3p, 3d, 4s, 4p, 5s, and 5p.  The closed shells 1s$^2$ + 1s$^2$
form the core of the molecule and no excitations from 1s$^2$ + 1s$^2$
will be allowed.  The 2s$^2$, 2p$^6$ and 3s$^2$ orbitals are valence
orbitals and single and double excitations from these orbitals occur.
Various covalent and ionic configurations are constructed by distributing
electrons from the optimized valence orbitals in all allowed ways over
the 3p, 3d, 4s, 4p, 5s, and 5p orbitals. In total, there are 1041
molecular configurations in our configuration interaction. We have
performed two kinds of nonrelativistic calculations. The first kind is
aimed at calculating the best possible short-range potential by first
perturbatively estimating the correlation energy of each molecular
configuration excited from the ground state configuration. If the
estimate falls below a threshold this configuration is not included in the
configuration interaction procedure. This truncation of the configurations
is necessary because inclusion of all configurations in the configuration
interaction procedure is not computationally possible.  The second kind
of calculation gives the best possible long-range potential by excluding
ionic configurations and switching off the exchange interaction in the
Hamiltonian in order to reduce the size of the matrix and to accelerate
the calculation. The two calculations are connected between 13 $a_0$ and
14 $a_0$, because at these internuclear separations the difference between
the total energy of the two kinds of calculations, and thus the exchange
energy, is less than 5\% of the binding energy. Figure~\ref{all_pot} shows
our ground state $^1\Sigma_g^+$ potential of Mg$_2$.  The dissociation
energy is D$_e$ = 410 cm$^{-1}$.

\section{Long-range potentials}\label{lr}

The long-range dispersion potential of the Mg$_2$ ground state has
attracted considerable attention over the past few decades. Mg$_2$ was the
first alkaline earth van der Waals molecule for which a high accuracy RKR
potential was obtained and thus allowing a comparison with, or extraction
of, the long-range potential.  Stwalley \cite{Stwalley70,Stwalley} and
Li and Stwalley \cite{Li} constructed a dispersion potential using the
form $V(R) = D_0 - C_6/R^6 - C_8/R^8$ from the frequency-dependent atomic
dipole polarizability and the RKR curve. The dipole-dipole dispersion
coefficient $C_6$ was determined from the atomic polarizability
while both $D_0$, the energy difference between the $\nu = 0, J = 0$
rovibrational level and the asymptotic energy, and dipole-quadrupole
dispersion coefficient $C_8$ were then obtained by fitting to the RKR
curve minus the dipole-dipole dispersion contribution. They find $D_0$=
404.1(0.5) cm$^{-1}$, $C_6$= 683(35) a.u.  (1 a.u. = 1 Hartree $a_0^6$,
1 Hartree = 4.359743 $\times 10^{-18}$ J), and $C_8$= 38(8)$\times
10^3$ a.u. (1 a.u. = 1 Hartree $a_0^8$).  In Ref.  \cite{Stanton}
the $C_6$ dispersion coefficient has been calculated based on atomic
coupled-cluster theory.  The result of this calculation is $C_6$ =
647.8 a.u..  Upper and lower bounds for the dispersion coefficients
have been obtained in Ref.~\cite{Standard} using a Pade approximation to
bound the atomic multipole polarizabilities.  Their ranges are $C_6$ =
630 a.u. - 638 a.u. and $C_8$ = 41100 a.u. - 43500 a.u..  In addition they
estimated the dipole-octupole coefficient, $C_{10}$, to be between 2730000
a.u. and 3040000 a.u.  (1 a.u. = 1 Hartree $a_0^{10}$).  Recently, Porsev
and Derevianko \cite{Porsev} have calculated the $C_6$ coefficient using
accurate theoretical and experimental atomic data. Different contributions
to the $C_6$ coefficient were obtained with different atomic relativistic
many-body electronic structure methods.  The dominant contribution to
this coefficient was found by combining configuration interaction and
many-body perturbation theory. The reported value of $C_6=$ 627(12)
a.u. agrees quite well with the bounds of Ref.~\cite{Standard}.
Finally, the dispersion coefficients can be evaluated from molecular
electronic structure calculations. They are extracted by fitting to the
long-range shape of the interaction potential. For example, the induced
dipole-dipole dispersion coefficient of Ref. \cite{Czuchaj} is about
a factor of 2 larger, while that of our MVB calculation is about 15\%
smaller, than that of Ref. \cite{Porsev}.

The long-range behavior of the RKR and IPA potentials, the dispersion
potential obtained from Ref. \cite{Porsev,Standard}, and our MVB
calculation are shown in Fig.~\ref{ground2}. The full line with filled
diamonds and full line are the RKR and IPA potentials, respectively. The
dotted line of the MVB calculation smoothly connects to the RKR at 13.53
$a_0$. Although this smooth connection is fortuitous, it suggests that
the RKR potential be extrapolated by the MVB potential. On the other
hand the long-range dispersion potential denoted by the dash-dotted line
using the $C_6$ coefficient of Ref. \cite{Porsev} and $C_8=42300$ a.u. and
$C_{10}=2885000$ a.u. of Ref. \cite{Standard} smoothly connects to the IPA
potential.  The $C_6$ coefficient of Ref.~\cite{Czuchaj} leads to a too
attractive long-range behavior as shown by the dashed curve in the figure.

For our study of scattering properties the spectroscopically derived RKR
and IPA potentials are extrapolated to large internuclear separations
$R$ using the dispersion plus exchange form $V_{dis}(R) + V_{ex}(R)$,
where $V_{dis}(R) = - \sum_{n=6,8,10}C_n/R^n$ and $V_{ex}(R) = B\times
R^{\alpha}\times exp(-\beta R)$.  For $R$ $>$ 13.53 $a_0$, the largest
turning point of the RKR potential, this potential is best extrapolated
by the dispersion coefficients extracted from the MVB potential and the
exchange potential of Ref.~\cite{Smirnov}. The exchange contribution
is small but has been added for completeness sake.  Other values for
these coefficients do not seem to be consistent with the shape of the
RKR potential. The IPA potential ending at $R$ = 14.5 $a_0$ is smoothly
joined to $V_{dis}(R)$ + $V_{ex}$(R) starting at $R$ = 16 $a_0$ The $C_6$
coefficient is taken from Ref.~\cite{Porsev} and the $C_8$ and $C_{10}$
coefficients are taken from Ref.~\cite{Standard}. The exchange potential
is from Ref.~\cite{Smirnov} where the exchange parameters are $\alpha$
= 3.63, $\beta$ = 1.512 1/$a_0$, and $B$ = 0.27 a.u., respectively.

\section{Vibrational bound states and scattering length}\label{vib}

\begin{figure} 
\psfig{file=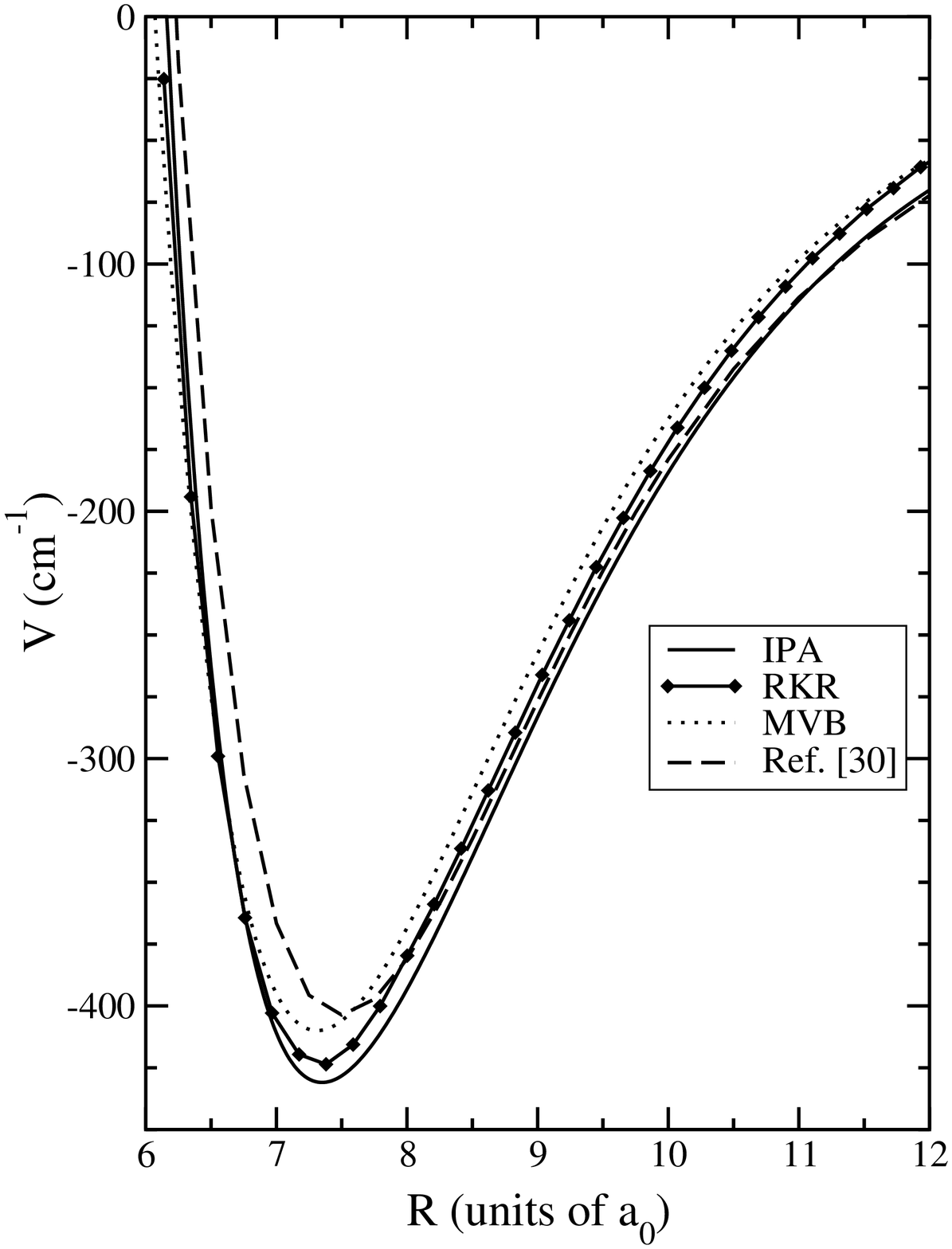,width=8.5cm,clip=} 
\caption{Ground state $^1\Sigma^+_g$ potential energy curves obtained
by different methods as a function of internuclear separation $R$. The
internuclear separation is in units of $a_0$ = 0.0529177 nm.  The solid
line with filled diamonds corresponds to the RKR potential; the solid
line shows the IPA potential; the dotted line is our MVB potential and
the dashed line is the potential by Ref.~{\protect\cite{Czuchaj}} }
\label{all_pot} 
\end{figure}

\begin{figure} 
\psfig{file=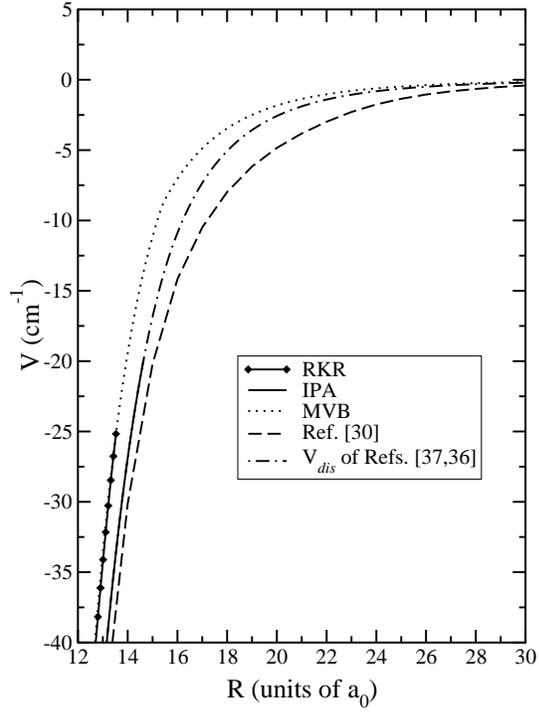,width=8.5cm,clip=} 
\caption{$^1\Sigma^+_g$ potential energy curves as a function of
internuclear separation. The region where the RKR and IPA potential
are connected to the long-range dispersion potential is shown. The line
style is the same as that used in Fig.~{\protect\ref{all_pot}} where in
addition the dash-dotted line is the dispersion potential using the data
from Refs.~{\protect\cite{Porsev,Standard}}.} 
\label{ground2} 
\end{figure}

In order to test the potentials constructed in the previous sections we
numerically calculated the rovibrational bound states and compared them
with the experimentally obtained term values of Ref.~\cite{Balfour}. The
eigenvalues of the Schr\"{o}dinger equation for the ground state
Hamiltonian have been obtained for each of the four (splined) adiabatic
potentials shown in Figs. \ref{all_pot} and \ref{ground2}. The ground
state Hamiltonian includes the electrostatic interaction in form of
the $^1\Sigma_g^+$ Born-Oppenheimer potential, the mechanical rotation
operator, $\hat{\it l}^2/2\mu R^2$, and the kinetic energy operator.

The experimental term values are relative to the $\nu = 0, J = {\it l}
= 0$ bound state.  Consequently, theoretical term values of a potential
are defined relative to the energy of the $\nu = 0, J = 0$ level of
this potential.  Our difference measure $\Delta$ is given by the square
root of the averaged squared difference between the theoretical and
experimental term values.  The difference averaged  over 254 ($\nu =
0 - 12$, J = {\it l} = 10 - 68) rovibrational levels lying below the
dissociation energy for four potentials is given in Table~\ref{deviation}.
The table shows that the IPA potential is one order of magnitude better
than the RKR potential in representing the experimental term values,
confirming the predicted improvement by Ref. \cite{Vidal}. The MVB
potential is about as accurate as the RKR potential while the potential
of Ref. \cite{Czuchaj} is clearly the least accurate.

We have set up a quantum scattering calculation for two ground state
Mg atoms.  For ultra-cold atom physics the relevant properties of
an interaction potential are the number of s-wave bound states and
the scattering length $a$ at zero collision energy. The number of
bound states of a potential is equal to the number of nodes of the
zero-energy scattering wavefunction. We find that the number of bound
states is different for each of our four potentials. In fact, we have
18, 19, 18, and 20 bound states for the RKR-, IPA-, and MVB-potential,
and the potential of Ref.~\cite{Czuchaj}, respectively. Qualitatively,
this variation can be understood from the long-range shape of these four
potentials. The dispersion potential of Ref.~\cite{Czuchaj} is the most
attractive of all four potentials consistent with the observation that it
has the largest number of bound states. The difference in dissociation
energy does not lead to a change in the number of bound states. Notice
that $v$=12 is the most-weakly-bound experimentally observed s-wave
vibrational level, that is, the last seven vibrational levels have not
been observed.

For the remainder of the paper we restrict scattering calculations
to the IPA potential, since the discrepancy with the experimental term
values for the IPA potential is an order of magnitude better than that of
the others.  The short-range part of the IPA potential smoothly connects
to the state-of-the-art long-range dispersion and exchange potential of
Refs.~\cite{Porsev,Standard,Smirnov} and thus gives added confidence in
the potential.

The IPA potential is tabulated in Table~\ref{potential} and when
splined has a scattering length of 26 $a_0$.  The scattering length of a
potential is defined by the phase of the wavefunction at zero collision
energy. It depends on the binding energy of the most-weakly bound $s$-wave
vibrational level.  In order to obtain the accuracy of the scattering
length for the IPA potential we must determine the effect of small
changes to the potential. The uncertainty in the dissociation energy,
the long-range dispersion coefficients, and the shape of the short-range
potential limit the accuracy of our determination.  The allowed variations
are chosen within the published uncertainties.  The 1 cm$^{-1}$
uncertainty of the dissociation energy of Ref.~\cite{Vidal} leads to
a 7 $a_0$ uncertainty in $a$. The effects of the uncertainty in the
dissociation energy was studied by uniformly shifting the IPA potential
for $R<$ 14.5 $a_0$, fitting smoothly to the dispersion potential.

The 12 a.u. uncertainty in the $C_6$ coefficient of Ref.~\cite{Porsev}
adds an additional 7 $a_0$ uncertainty to $a$, while uncertainties
in the $C_8$ and $C_{10}$ coefficients add 1.4 $a_0$ and 0.5 $a_0$,
respectively. These corrections to the potential do not change the
agreement between the calculated and experimental bound state term values.
Nevertheless, a $\Delta$ of 0.22 cm$^{-1}$ does imply that the shape
of potential is not fully characterized, which introduces additional
uncertainties to the scattering length. Local changes to the potential
on the order of $\Delta$ are needed.  Since $\Delta$ is about six times
smaller than the uncertainty in the dissociation energy we can estimate
the additional uncertainty in $a$ to be 1 $a_0$.  The largest contributions
to the uncertainty in the scattering length are, therefore, due to the
uncertainty in $D_e$ and $C_6$.  None of the discussed uncertainties in
the potential changes the number of bound states.  The final value for
the scattering length of the IPA potential is 26(10) $a_0$ or 1.4(5) nm
by adding the uncertainties in quadrature. The potential has 19 $s$-wave
bound states.

\begin{figure} 
\psfig{file=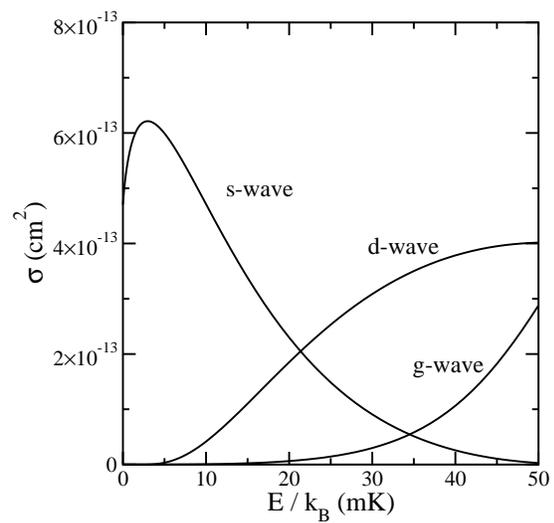,width=8.5cm,clip=}
\caption{The $s$, $d$, and $g$-wave cross-section of the ground state Mg
+ Mg collision as a function of collision energy.} 
\label{cross-section}
\end{figure}

Figure~\ref{cross-section} shows partial cross sections of two
colliding ground state Mg atoms as a function of collision energy.
The contributions of $s$-, $d$-, and $g$-wave collisions are shown.
The partial-wave cross section is defined as 
\begin{eqnarray}
   \sigma_l &=& (2l+1) \frac{8\pi}{k^2}\sin^2\delta_l(k)
\end{eqnarray} 
where $\delta_l(k)$ is the $^1\Sigma_g^+$ phase shift for $l$-wave
collisions and $k$ is the relative wavenumber of the collision.
The cross section is calculated for the potential which has $a$=26 $a_0$
and is tabulated in Table~\ref{potential}. The $s$-wave cross section
has a maximum at $E/k_B$= 3 mK and then decreases monotonically up to
$E/k_B$=50 mK. At zero collision energy the $s$-wave cross section equals
$8\pi a^2$.  The $d$- and $g$-wave partial cross sections are zero at
zero collision energy and then rise rapidly and have a maximum at much
higher collision energy than that given by the height of the $d$- and
$g$-wave centrifugal barrier.  In fact, the centrifugal barriers are at
$E/k_B$= 7.8 mK and 47 mK for the $d$- and $g$-wave, respectively.

In summary, we have constructed four $^1\Sigma_g^+$ potentials for the
collision between two ground state Mg atoms. Two of these potentials
are based on the experimentally determined term values of the dimer,
while the other two were of a theoretical origin. The MVB theoretical
potential is calculated in this Paper.  By comparing the experimental term
values with those calculated for the four Born-Oppenheimer potentials we
conclude that the theoretical MVB potential is of similar accuracy as
the ``experimentally determined'' RKR potential. Given the difficulty
of introducing electron correlations into the molecule the agreement
is remarkable.  Existing long-range dispersion and exchange coefficients
have been discussed as well.

The best $^1\Sigma_g^+$ potential is obtained from an IPA potential
for $R< 14.5$ $a_0$ connected to a dispersion potential based on
Refs.~\cite{Porsev} and \cite{Standard}. This potential reproduces
the experimental term values to 0.2 cm$^{-1}$, an order of magnitude
better than the comparison for the other three potentials, and gives a
scattering length of +1.4(5) nm where the error is prodominantly due to
the uncertainty in $D_e$ and the $C_6$ dispersion coefficient. A positive
sign of the scattering length suggests that a Bose-Einstein condensate
of ground state Mg atoms is stable.

\begin{table} 
\caption{The square root of the averaged squared
difference between four calculated Born-Oppenheimer potentials of
the X$^1\Sigma^+_g$ state and measured rovibrational energies of
Ref.~{\protect\cite{Balfour}}.} 
\label{deviation} 
\begin{center}
\begin{tabular}{lc} 
Potential  & Difference $\Delta$(cm$^{-1}$) \\
\hline RKR            &     2.1 \\ IPA            &     0.2 \\ {\it
Ab~initio} Ref.~\cite{Czuchaj} &  11.9   \\ {\it Ab~initio} this work &
2.3 
\end{tabular} 
\end{center} 
\end{table}

\begin{table} 
\caption{Constructed IPA potential for the X$^1\Sigma^+_g$
ground potential of Mg$_2$. For $R>18$ $a_0$ the analytical dispersion
and exchange potential discussed in the text is used.} 
\label{potential}
\begin{tabular} {lrlr} R($a_0$) &V(cm$^{-1}$)&  R($a_0$)&  V(cm$^{-1}$)\\
\hline 5.00 &  3684.419 & 8.50 & -339.579\\ 5.10 &  2989.552 & 8.70 &
-316.957\\ 5.20 &  2409.684 & 8.90 & -294.292\\ 5.30 &  1925.473 & 9.10 &
-272.121\\ 5.40 &  1520.991 & 9.30 & -250.742\\ 5.50 &  1183.093 & 9.50 &
-230.345\\ 5.60 &  900.903 & 9.70 & -211.057\\ 5.70 &  665.394 & 9.90 &
-192.949\\ 5.80 &  469.063 & 10.10 & -176.055\\ 5.90 &  305.652 & 10.30 &
-160.376\\ 6.00 &  169.935 & 10.50 & -145.892\\ 6.10 &  57.539 & 10.70 &
-132.563\\ 6.20 &  -39.100 & 10.90 & -120.340\\ 6.30 &  -130.963 & 11.10 &
-109.165\\ 6.40 &  -204.311 & 11.30 & -98.974\\ 6.50 &  -262.877 & 11.50 &
-89.703\\ 6.60 &  -309.439 & 11.70 & -81.285\\ 6.70 &  -346.071 & 11.90 &
-73.654\\ 6.80 &  -374.322 & 12.10 & -66.749\\ 6.90 &  -395.542 & 12.30 &
-60.505\\ 7.00 &  -411.247 & 12.50 & -54.867\\ 7.06 &  -418.550 & 12.70 &
-49.782\\ 7.14 &  -425.235 & 12.90 & -45.205\\ 7.22 &  -428.791 & 13.10 &
-41.098\\ 7.30 &  -430.650 & 13.30 & -37.398\\ 7.36 &  -430.956 & 13.50 &
-34.043\\ 7.44 &  -429.909 & 13.70 & -30.982\\ 7.50 &  -428.008 & 13.90 &
-28.190\\ 7.56 &  -425.243 & 14.10 & -25.649\\ 7.62 &  -421.888 & 14.30 &
-23.341\\ 7.68 &  -418.145 & 14.50 & -21.248\\ 7.74 &  -414.151 & 15.00 &
-16.855\\ 7.80 &  -410.024 & 15.50 & -13.472\\ 7.86 &  -405.861 & 16.00 &
-10.896\\ 7.92 &  -401.352 & 16.50 & -8.910\\ 7.98 &  -396.181 & 17.00 &
-7.334\\ 8.10 &  -384.374 & 17.50 & -6.075\\ 8.30 &  -362.286 & 18.00 &
-5.062 
\end{tabular} 
\end{table}

\end{document}